\documentclass[preprint,showpacs,preprintnumbers,amsmath,amssymb,endfloats*]{revtex4}
\usepackage{graphicx}

\begin{document}

\thispagestyle{empty}
\title{Experimental approaches to the difference in
the Casimir force
through the varying optical properties of boundary surface
}

\author{R.~Castillo-Garza${}^{1}$, C.-C.~Chang${}^{1}$,
D.~Jimenez${}^{1}$, G.~L.~Klimchitskaya${}^2$,
V.~M.~Mostepanenko${}^3$, and U.~Mohideen${}^1$}

\affiliation{${}^{1}$Department of Physics and Astronomy,
University of California,
Riverside, California 92521, USA \\
${}^2$North-West Technical
University, Millionnaya St. 5, St.Petersburg 191065, Russia\\
${}^3$Noncommercial Partnership ``Scientific Instruments'',
Tverskaya St.{\ }11, Moscow 103905, Russia}

\begin{abstract}
We propose two novel experiments on the measurement of the
Casimir force acting between a gold coated sphere and
semiconductor plates with markedly different charge carrier
densities. In the first of these experiments a patterned Si
plate is used which consists of two sections of different dopant
densities and oscillates in the horizontal direction below
a sphere. The measurement scheme in this experiment is
differential, i.e., allows the direct high-precision
measurement of the difference of the Casimir forces between
the sphere and sections of the patterned plate or the difference
of the equivalent pressures between Au and patterned
parallel plates with static and dynamic techniques, respectively.
The second experiment proposes to measure the Casimir force
between the same sphere and a VO${}_2$ film which undergoes
the insulator-metal phase transition with the increase of
temperature. We report the present status of the interferometer
based variable temperature apparatus developed to perform
both experiments and present the first results on the
calibration and sensitivity. The magnitudes of the Casimir forces
and pressures in the experimental configurations are calculated
using different theoretical approaches to the description of
optical and conductivity properties of semiconductors at low
frequencies proposed in the literature. It is shown that the suggested
experiments will aid in the resolution of theoretical problems
arising in the application of the Lifshitz theory at nonzero
temperature to real materials. They will also open new
opportunities in nanotechnology.
\end{abstract}


\maketitle

\section{Introduction}

The Casimir effect \cite{1} implies that there is a force acting
between closely spaced electrically neutral bodies following from
the zero-point oscillations of the electromagnetic field.
The Casimir force can be viewed as an extension of the van der
Waals force to large separations where the retardation effects
come into play. Within a decade of Casimir's work, Lifshitz
and collaborators \cite{2,3} introduced the role of
optical properties of the material
into the van der Waals and Casimir force.
In the last few years, the advances following from both
fundamental physics and nanotechnology have motivated careful
experimental and theoretical investigations of the Casimir
effect. The first modern experiments were made with metal test
bodies in a sphere-plate configuration, and their results are
summarized in Ref.~\cite{4}. In subsequent experiments the
lateral Casimir force between corrugated surfaces \cite{5}
and the pressure in the original Casimir configuration
\cite{6} have been demonstrated. Later experiments \cite{7,8,8a}
have brought the most precise determination of the Casimir
pressure between two metal plates. The rapid theoretical
progress has raised fundamental questions on our
understanding of the Casimir force between real metals at
nonzero temperature. Specifically, the role of conductivity
processes and the related optical properties of metals at
quasi-static frequencies has become the subject of
discussions \cite{9,10,11,12,13,14,15,16}.

One of the most important applications of the Casimir effect
is the design, fabrication and function of micro- and
nanoelectromechanical systems such as micromirrors,
microresonators, nanotweezers and nanoscale actuators
\cite{17,18,19,20,21}. The separations between the adjacent
surfaces in such devices are rapidly falling below a micrometer,
i.e., to a region where the Casimir force becomes comparable
with typical electrostatic forces. It is important that
investigations of the Casimir force be done in semiconductors
as they are the material of choice for the fabrications of
optomechanical, micro- and nanoelectromechanical systems.
While the role of conductivity and optical properties of
materials can be checked in metals, semiconductors offer
better control of the related parameters (charge carrier
density, defect density, size etc.) and will provide an
exhaustive check of the various models.

Reference \cite{22} pioneered the measurement of the Casimir
force acting between a semiconductor surface, single
crystal Si wafer, and a gold coated sphere. The experimental
data obtained for a wafer with the concentration of charge
carriers $\approx 3\times 10^{19}\,\mbox{cm}^{-3}$ were
compared with the Lifshitz theory at zero temperature and
good agreement was found at a 95\% confidence level.
At the same time, the theory describing a ``dielectric''
Si plate with a concentration
$\sim 5\times 10^{12}\,\mbox{cm}^{-3}$ was excluded by
experiment at 70\% confidence. This allows one to
conclude that the Casimir force is sensitive to the
conductivity properties of semiconductors. This conclusion
has found direct experimental confirmation in Ref.~\cite{23}
where the Casimir forces between a gold coated sphere and
two different Si wafers with the concentrations of charge
carriers $\approx 3.2\times 10^{20}\,\mbox{cm}^{-3}$ and
$1.2\times 10^{16}\,\mbox{cm}^{-3}$ have been measured.
The difference of the mesured forces for the two
conductivities was found to be in good agreement with the
corresponding difference of the theoretical results
computed at zero temperature (note that the sensitivity of
force mesurements in Refs.~\cite{22,23} was not sufficient
to detect the thermal corrections predicted in
Refs.~\cite{9,10,11,12,13,14,15,16}).

In the most precise experiment on the Casimir force between
a metal and a semiconductor \cite{24}, the density of charge
carriers in a Si membrane of 4$\,\mu$m thickness is changed from
$5\times 10^{14}\,\mbox{cm}^{-3}$ to
$2\times 10^{19}\,\mbox{cm}^{-3}$
through the absorption of photons from a laser pulse.
This is a differential experiment where only the difference
of the Casimir forces in the presence and in the absence
of a laser pulse was measured. This decreases the
experimental error to a fraction of a 1\,pN and allows one to
check the role of conductivity processes in semiconductors
at the laboratory temperature $T=300\,$K. The experimental
data for the difference Casimir force as a function
of separation were compared with the Lifshitz theory and
the outcome was somewhat puzzling. The data were found
to be in excellent agreement with the theoretical
difference force computed at $T=300\,$K under the assumption
that in the absence of laser light Si possesses a finite
static dielectric permittivity. By contrast, if theory
takes into account the dc conductivity of Si in the absence
of laser light, it is excluded by the data at a 95\% confidence
level. This is somewhat analogous to the above-mentioned
problems for two real metals where the inclusion of the actual
conductivity processes at low frequencies also leads to
disagreement with experiment \cite{7,8,8a}. The fundamental
questions on the role of scattering processes and conductivity
at low frequencies in the Casimir force have to be clarified
for further progress in the field and this calls for new
precise experiments.

In this paper we propose two experiments on the Casimir
force between a metal sphere and semiconductor plate which
can shed light on the applicability of the Lifshitz theory
at nonzero temperature to real materials. In the first of
these experiments, the patterned Si plate with two sections
of different dopant densities is oscillated in the
horizontal direction below the Au coated sphere. As a result,
the sphere is subject to the difference Casimir force which
can be measured using the static and dynamic techniques.
This experimental scheme promises a record sensitivity
to force differences at the level of 1\,fN. In the second
experiment, we propose to demonstrate the modulation of the
Casimir force by optically switching the
insulator-metal transition in VO${}_2$ films \cite{25}.
The phase transition between the insulator and metal leads
to a change in the charge carrier density of order
$10^4$, which is sufficient to bring about a large change
in the Casimir force. For both experiments the related
theory is elaborated and the magnitudes of Casimir forces
are computed with application to the experimental configurations.
The effects from using different theoretical approaches
to the description of conductivity processes are
carefully analyzed and shown to be observable in the
proposed experiments. The present
status of developing the apparatus at UC Riverside, its
calibration and sensitivity is presented. The proposed
experiments offer a precision test of the role of
conductivity, optical properties and scattering in the
Lifshitz theory of the van der Waals and Casimir force
at nonzero temperature. They also open up possibilities
of radically new nanomechanical devices using the
Casimir force in imaging applications.

The paper is organized as follows. In Sec.~II, we describe
the proposed experiment on the difference Casimir force
with the patterned semiconductor plate. The brief description
of the experimental apparatus and preliminary results are also provided.
Section~III contains the calculation of the difference
Casimir force and equivalent pressure in the patterned geometry
using the Lifshitz theory at nonzero temperature and different
models of conductivity processes at low frequencies.
In Sec.~IV we propose the experiment on the modulation of the
Casimir force through a metal-insulator transition. The
experimental scheme and some preliminary tests are discussed.
Section~V presents theoretical computations of Casimir forces
in a insulator-metal transition on the basis of the Lifshitz
theory at nonzero temperature and using different models for the
conductivity processes. Section VI contains our conclusions and
a discussion.

\section{Proposed experiment on the difference Casimir force
with the patterned semiconductor plate}

The aim of this experiment is to gain a fundamental understanding
of the role of carrier density in the Casimir force using a
nanofabricated patterned semiconductor plate.
The proposed design for the experiment is shown schematically
in Fig.~1. The gold coated polystyrene sphere of about 100$\,\mu$m
radius is attached to a cantilever of an atomic force microscope
(AFM) specially adapted for making sensitive force measurements.
Instead of the simple single crystal Si substrate used in
the previous experiments \cite{22,23}, here a patterned Si
plate is employed. This plate is composed of single crystal Si
specifically fabricated to have adjacent sections of two different
charge carrier densities $n\sim 10^{16}\,\mbox{cm}^{-3}$ and
$\tilde{n}\sim 10^{20}\,\mbox{cm}^{-3}$. In this range of doping
densities, the plasma frequency $\omega_p$ will change by a
factor of 100. Additional changes in the $\omega_p$ can be
brought about by using both $p$(B) and $n$(P) type dopants as
electrons and holes differ in their effective mass by 30\%.

The preparation of the Si sample with the two sections having
different conductivities is shown schematically in Fig.~2.  First, one
half of the bare Si wafer of 0.3 to 0.5\,mm
thickness \cite{22,23} having the lower conductivity shown in (a)
is masked with a photoresist as shown in (b).  Next in (c) the
exposed half of the Si wafer is doped with P ions using ion
implantation leading to a higher density of electrons in the
exposed half in (d).
Rapid thermal annealing and chemical mechanical
polishing of the patterned Si plate will be done as the last step.
This is to ensure that there are no surface features resulting
from the fabrication. Similar nanofabrication procedures of
semiconductors were used in our previous work \cite{23,26,27}.
Sharp transition boundaries between the two sections of Si plate
of width less than 200\,nm are possible. The limitation comes from
interdiffusion and the resolution of the ion implantation
procedure to be used. It might be necessary to further limit
interdiffusion by the creation of a narrow 100\,nm barrier
between the two doped regions.

Identically prepared but unpatterned samples will be used to
measure the properties which are needed for theoretical
computations. The carrier concentration will be measured using
Hall probes. This will yield an independent measurement of the
plasma frequencies. A four probe technique will be used to
measure the conductivity $\sigma$. From the conductivity and
the charge carrier concentration, the scattering time
$\tau=\sigma/(\varepsilon_0\omega_p^2)$ will be found,
where $\varepsilon_0$ is the permittivity of free space.

The first important improvement of this experiment, as compared
with Refs.~\cite{22,23}, is the direct measurement of
difference forces when the patterned plate is oscillated
below the sphere. This measurement is performed as follows.
The patterned Si plate will be mounted on the piezo below a Au
coated sphere as is shown in Fig.~1. The Si plate is
positioned such that the boundary is below the vertical diameter of
the sphere. The distance between the sphere and Si plate $z$ will
be kept fixed and the Si plate will be oscillated in the
horizontal direction using the piezo such that the sphere crosses
the boundary in the perpendicular direction during each oscillation
(a similar approach was exploited in Ref.~\cite{28} for
constraining new forces from the oscillations
of the Au coated sphere
over two dissimilar metals, Au and Ge). The Casimir force on
the sphere changes as the sphere crosses the boundary. This
change corresponds to the differential force
\begin{equation}
\Delta F(z)=F_{\tilde{n}}(z)-F_{n}(z),
\label{eq1}
\end{equation}
\noindent
equal to the difference of the Casimir forces due to two
different charge carriers densities $\tilde{n}$ and $n$,
respectively. This causes a difference in the deflection
of the cantilever.
In order to reduce the random noise by averaging, the
periodic horizontal movement of the plate will be of
an angular frequency $\Omega\sim 0.1\,$Hz. The amplitude
of the plate oscillations is limited by the piezo
characteristics, but will be of order 100\,$\mu$m, much
larger than typical transition region of 200\,nm. The experiment
will be repeated for different sphere-plate separations in the
region from 100 to 300\,nm. The measurement of absolute
separations will be performed by the application of voltages
to the test bodies as described in Ref.~\cite{22}.

The second major improvement in this experiment in comparison
with all previous measurements of the Casimir force is the
increased sensitivity. This will be achieved through the use
of the interferometer based low temperature AFM capable of
operating over wide temperature range spanning from 360 to
4\,K, and the use of two measurement techniques, a static one and
a dynamic one. A picture of the newly constructed
experimental apparatus of a low-temperature AFM is shown
in Fig.~3. Here the cantilever deflection is measured
interferometrically and therefore has much higher sensitivity
than photodiodes used in the previous work \cite{22,23,24}.
The detection of a difference force $\Delta F(z)$ will be
done by two alternative techniques. The first technique,
a static one, reduces to the direct measurement of $\Delta F(z)$
as described above. We are presently performing the initial
tests and calibration trials at 77\,K. An oil free vacuum with
a pressure of around $2\times 10^{-7}\,$Torr is used. The
instrument is magnetically damped to yield low mechanical
coupling to the environment. The temperature can be varied
with a precision of 0.2\,K. We have fabricated special
conductive cantilevers with a spring constant $k=0.03\,$N/m.
The magnitude of $k$ is found by applying electrostatic
voltages to the plate as discussed in Ref.~\cite{22}.
To accomplish this, Si cantilevers were thermal diffusion
doped to achieve the necessary conductivity. Note that
conductive cantilevers are necessary to reduce electrostatic
effects. The cantilever-sphere arrangement has been
checked to be stable at 77\,K.

The experimental setup using the static
measurement technique allows not
only a demonstration but a detailed investigation of
the influence of carrier density, conductivity and scattering in
semiconductors on the Casimir force. According to the
results of Refs.~\cite{22,23,24} and calculations below in
Sec.~III, the magnitude of the difference force to be measured
is about several pN. Different theoretical models of
conductivity processes at low frequencies lead to predictions
differing by approximately 1\,pN within a wide separation
region (see Fig.~5 in Sec.~III). The described setup provides
excellent opportunity for precise measurement of forces of order
and below 1\,pN. We have measured a resonance frequency of the
cantilever of $f_r=1130.9\,$Hz, a quality factor $Q=5889.2$,
and an equivalent noise bandwidth $B=0.3\,$Hz. The resultant
force sensitivity of our cantilever at $T=77\,$K with the
gold coated sphere attached was determined following
Ref.~\cite{29} to be
\begin{equation}
\delta F_{\min}=\left(
\frac{2k_B TkB}{\pi Qf_r}\right)^{1/2}\approx
0.96\times 10^{-15}\,\mbox{N}\approx
1\,\mbox{fN},
\label{eq2}
\end{equation}
\noindent
where $k_B$ is the Boltzmann constant. Even bearing in mind
that the systematic error may be up to an order of magnitude
larger, the sensitivity (\ref{eq2}) presents considerable
possibilities for the precise investigation of the
difference Casimir force $\Delta F$.

The second technique for the detection of the difference Casimir
force is a dynamic one \cite{7,8,20}. Here this technique is
applied not for a direct measurement of $\Delta F$ but rather
for the experimental determination of the equivalent difference
Casimir pressure between the two parallel plates (one made
of Au and the other one, a patterned Si plate).
The cantilever-sphere system oscillates
in the verical direction due its thermal noise
with a resonant frequency $\omega_r=(k/M)^{1/2}=2\pi f_r$
in the absence of the Casimir force, where $M$ is the mass of
the system. The thermal noise spectrum of the sphere-cantilever
system is measured and fit to a Lorentzian to identify the
peak resonant frequency, $\omega_r$.
The shift of
$\omega_r$ in the presence of the Casimir force
when  for example the sphere is positioned above a section
of the patterned Si plate with the density
of charge carriers $\tilde{n}$
is  equal to \cite{7,8,20}
\begin{equation}
\omega_{r,\tilde{n}}-\omega_r=-\frac{\omega_r}{2k}\,
\frac{\partial F_{\tilde{n}}(z)}{\partial z}.
\label{eq3}
\end{equation}

Next the  plate is oscillated
in the horizontal direction with a
frequency $\Omega$. As a result, the frequency shift
\begin{equation}
\omega_{r,\tilde{n}}-\omega_{r,n}=-\frac{\omega_r}{2k}\,
\frac{\partial\Delta F(z)}{\partial z}
\label{eq4}
\end{equation}
\noindent
between the resonant frequencies above the two different sections of
the patterned Si plate is measured.
Using the proximity force approximation
\cite{4,30}, we determine the difference Casimir pressure
\begin{equation}
\Delta P(z)=-\frac{1}{2\pi R}\,
\frac{\partial\Delta F(z)}{\partial z}
\label{eq5}
\end{equation}
\noindent
between the two parallel plates (the Au one and the patterned
Si). Note that the systematic error from the use of the
proximity force approximation was recently confirmed to be less
than $z/R$ \cite{31,32,33,34,34a}. Equations (\ref{eq4}) and
(\ref{eq5}) express the difference Casimir pressure through the
measured shift of the resonance frequency above the two halfs of
the patterned plate. As is shown in the next section, the
measurements of the difference Casimir pressure using the dynamic
technique provides us with one more test of the predictions of the
different models of conductivity processes at low frequencies.

The experiment on the difference Casimir force from a patterned
Si plate, as described in this section, allows variation of
charge carrier density by the preparation of different
semiconductor samples. Thus, the proposed measurements should
provide a comprehensive understanding on the role of conductivity
 and optical processes in the Casimir force for nonmetallic
materials and discriminate between competing theoretical
approaches.

\section{Calculation of the difference Casimir force and
\protect{\\} pressure in the patterned geometry}

The difference Casimir force and the equivalent Casimir
pressure from the oscillation of the patterned Si plate
below an Au coated sphere at $T=300\,$K in thermal equilibrium are
given by the Lifshitz theory. In the static technique the data to be
compared with theory is the difference of Casimir forces
acting between the sphere and two sections of the patterned plate.
This difference is obtained from the Casimir energy between two
parallel plates, as given by the Lifshitz theory, using the
proximity force approximation \cite{2,3,4}
\begin{eqnarray}
&&
\Delta F(z)=k_BTR\sum\limits_{l=0}^{\infty}\left(1-
\frac{1}{2}\delta_{l0}\right)\int_{0}^{\infty}
k_{\bot}dk_{\bot}
\label{eq6} \\
&&\phantom{aa}\times
\ln\frac{\left[1-r_{TM;\tilde{n}}(\xi_l,k_{\bot})
r_{TM}(\xi_l,k_{\bot})e^{-2q_lz}\right]\,\left[1-
r_{TE;\tilde{n}}(\xi_l,k_{\bot})
r_{TE}(\xi_l,k_{\bot})e^{-2q_lz}\right]}{\left[1-
r_{TM;{n}}(\xi_l,k_{\bot})
r_{TM}(\xi_l,k_{\bot})e^{-2q_lz}\right]\,\left[1-
r_{TE;{n}}(\xi_l,k_{\bot})
r_{TE}(\xi_l,k_{\bot})e^{-2q_lz}\right]}\,.
\nonumber
\end{eqnarray}
\noindent
Here $\xi_l=2\pi k_BTl/\hbar$ with $l=0,\,1,\,2,\,\ldots$
are the Matsubara frequencies,
$q_l=(k_{\bot}^2+\xi_l^2/c^2)^{1/2}$, and $k_{\bot}$ is the
projection of the wave vector on the boundary planes. The reflection
coefficients on the Au plane for the two independent polarizations
of the electromagnetic field (transverse magnetic and transverse
electric modes) are
\begin{equation}
r_{TM}(\xi_l,k_{\bot})=
\frac{\varepsilon_lq_l-k_l}{\varepsilon_lq_l+k_l},
\qquad
r_{TE}(\xi_l,k_{\bot})=
\frac{k_l-q_l}{k_l+q_l},
\label{eq7}
\end{equation}
\noindent
where $k_l=(k_{\bot}^2+\varepsilon_l\xi_l^2/c^2)^{1/2}$ and
$\varepsilon_l=\varepsilon(i\xi_l)$ is the dielectric
permittivity of Au along the imaginary frequency axis.
In a similar way, the reflection coefficients on the
two sections of a
patterned Si plate with charge carrier densities $\tilde{n}$ and
$n$ are given, respectively, by
\begin{equation}
r_{TM;\tilde{n},n}(\xi_l,k_{\bot})=
\frac{\varepsilon_{l;\tilde{n},n}q_l-
k_{l;\tilde{n},n}}{\varepsilon_{l;\tilde{n},n}q_l+k_{l;\tilde{n},n}},
\qquad
r_{TE;\tilde{n},n}(\xi_l,k_{\bot})=
\frac{k_{l;\tilde{n},n}-q_l}{k_{l;\tilde{n},n}+q_l},
\label{eq8}
\end{equation}
\noindent
where
$k_{l;\tilde{n},n}=
(k_{\bot}^2+\varepsilon_{l;\tilde{n},n}\xi_l^2/c^2)^{1/2}$ and
$\varepsilon_{l;\tilde{n},n}=\varepsilon_{\tilde{n},n}(i\xi_l)$
are the dielectric permittivities of Si with charge carrier
densities $\tilde{n}$ and $n$ along the imaginary frequency axis.

In the dynamic technique the data to be compared with theory is
the equivalent difference Casimir pressure between two
parallel plates, one made of Au and the other one a patterned Si
plate. Using the same notations as above, the difference Casimir
pressure is given by
\begin{eqnarray}
&&
\Delta P(z)=-\frac{k_BT}{\pi}\sum\limits_{l=0}^{\infty}\left(1-
\frac{1}{2}\delta_{l0}\right)\int_{0}^{\infty}
k_{\bot}dk_{\bot}q_l
\label{eq9} \\
&&\phantom{aa}\times
\left\{\left[r_{TM;\tilde{n}}^{-1}(\xi_l,k_{\bot})
r_{TM}^{-1}(\xi_l,k_{\bot})e^{2q_lz}-1\right]^{-1}+\left[
r_{TE;\tilde{n}}^{-1}(\xi_l,k_{\bot})
r_{TE}^{-1}(\xi_l,k_{\bot})e^{2q_lz}-1\right]^{-1}\right.
\nonumber \\
&&
\phantom{aaaa}\left.
-\left[r_{TM;{n}}^{-1}(\xi_l,k_{\bot})
r_{TM}^{-1}(\xi_l,k_{\bot})e^{2q_lz}-1\right]^{-1}-
\left[r_{TE;{n}}^{-1}(\xi_l,k_{\bot})
r_{TE}^{-1}(\xi_l,k_{\bot})e^{2q_lz}-1\right]^{-1}\right\}\,.
\nonumber
\end{eqnarray}
\noindent
Note that in Eqs.~(\ref{eq6}) and (\ref{eq9}) we have replaced the
100\,nm Au coating and the 0.3--0.5\,mm Si plate for an Au and Si
semispaces, respectively. Using the Lifshitz formula for layered
structures \cite{4} it is easy to calculate the force and pressure
errors due to this replacement. For example, for an Au layer at a
typical separation of 100\,nm this error is about 0.01\%.
For Si a finite thickness of the plate $d$ markedly affects
the Casimir force when the separation distance $z$ exceeds the thickness,
i.e., $z/d>1$ \cite{35a}. In our case, however, even at the largest
separation considered ($z=300\,$nm) the ratio of the separtion
to the plate thickness $z/d\leq 10^{-3}$. This is similar to the
case of the experiment \cite{24} where
the finite thickness of Si membrane also does not influence the
magnitude of the Casimir force because at
separations $z\leq 200\,$nm where statistically meaningful results
were obtained $z/d\leq0.05$.

We have performed computations of the difference Casimir force
(\ref{eq6}) and difference Casimir pressure (\ref{eq9}) for
samples with typical values of charge carrier concentrations
$\tilde{n}$ and $n$ as used in experiments \cite{22,23,24}.
Both sections of the Si plate were chosen to have electron
conductivity and doped with P. For the section of the plate
with higher concentration of charge carriers the values
$\tilde{n}_1=3.2\times 10^{20}\,\mbox{cm}^{-3}$ (such a sample
was fabricated in Ref.~\cite{23}) and
$\tilde{n}_2=3.2\times 10^{19}\,\mbox{cm}^{-3}$ were used
in the computations. The respective dielectric permittivity along
the imaginary frequency axis can be represented in the form
\cite{34b}
\begin{equation}
\varepsilon_{\tilde{n}}(i\xi_l)=\varepsilon^{Si}(i\xi_l)+
\frac{\omega_{p;\tilde{n}}^2}{\xi_l(\xi_l+\gamma_{\tilde{n}})}.
\label{eq10}
\end{equation}
\noindent
Here $\varepsilon^{Si}(i\xi_l)$ is the permittivity of
high-resistivity (dielectric) Si along the imaginary frequency
axis computed in Ref.~\cite{35} by means of the dispersion
relation using the tabulated optical data for the complex index
of refraction \cite{36}. The values of the plasma frequencies and
relaxation parameters are the following \cite{23}:
$\omega_{p;{\tilde{n}_1}}=2.0\times 10^{15}\,$rad/s,
$\gamma_{{\tilde{n}_1}}=2.4\times 10^{14}\,$rad/s,
$\omega_{p;{\tilde{n}_2}}=6.3\times 10^{14}\,$rad/s,
$\gamma_{{\tilde{n}_2}}=1.8\times 10^{13}\,$rad/s.
In Fig.~4 the dielectric permittivities of the samples with high
concentrations of charge carriers $\tilde{n}_1$ and $\tilde{n}_2$
are shown as solid lines 1 and 2, respectively. In the same figure,
the dashed line $a$ shows the permittivity of high-resistivity,
dielectric, Si \cite{35} and the dotted line the permittivity
of Au computed in Ref.~\cite{35} using the tabulated optical data
of Ref.~\cite{36}.

Below we will use two models for the permittivity of the section
of the Si plate with lower concentration of charge carriers $n$.
Calculations show that for any
$0<n\leq 1.0\times 10^{17}\,\mbox{cm}^{-3}$ (this interval
includes the experimental value of
$n\approx 1.2\times 10^{16}\,\mbox{cm}^{-3}$ in Ref.~\cite{23}),
the obtained values of $F_n(z)$ and, thus, of $\Delta F(z)$
do not depend on $n$. Because of this we use in the computations
$n= 1.0\times 10^{17}\,\mbox{cm}^{-3}$,
the plasma frequency $\omega_{p;n}=3.5\times 10^{13}\,$rad/s
and the relaxation parameter $\gamma_{n}=1.8\times 10^{13}\,$rad/s
\cite{23,37} (note that for $n\leq 1.0\times 10^{17}\,\mbox{cm}^{-3}$
the value of the relaxation parameter does not effect the
magnitude of the Casimir force). Then the dielectric permittivity
of this section of the
Si plate along the imaginary frequency axis is given by
\begin{equation}
\varepsilon_{{n}}^{(b)}(i\xi_l)=\varepsilon^{Si}(i\xi_l)+
\frac{\omega_{p;{n}}^2}{\xi_l(\xi_l+\gamma_{{n}})}
\label{eq11}
\end{equation}
\noindent
and is shown as the dashed line $b$ in Fig.~4. This is one model
of Si with a lower concentration of charge carriers
referred to below as model ($b$).

As is seen in Fig.~4, the dashed line $b$, and thereby all respective
lines for samples with the concentration of charge carriers smaller
than $1.0\times 10^{17}\,\mbox{cm}^{-3}$, deviate from the
permittivity of dielectric Si (line $a$) only at frequencies below
the first Matsubara frequency $\xi_1$. Because of this, it is
common (see, e.g., \cite{2,3,38}) to neglect the small
conductivity of high-resistivity materials at low frequencies and
describe them in the frequency region below the first Matsubara
frequency by the static dielectric permittivity. In our case this
leads to
\begin{equation}
\varepsilon_{{n}}^{(a)}(i\xi_l)=\varepsilon^{Si}(i\xi_l),
\label{eq12}
\end{equation}
\noindent
which is the other model for Si with a lower concentration of charge
carriers referred to below as model ($a$).
From Eq.~(\ref{eq12}) and Fig.~4 at all frequencies
$\xi\leq\xi_1$ it follows:
$\varepsilon_{{n}}^{(a)}(i\xi)=\varepsilon^{Si}(0)=11.66$.

To be exact, at any $T>0$ the density of free charge carriers
$n$ in semiconductors (and even in dielectrics) and thus the conductivity
are nonzero ($n>0$).
Thus, the model (\ref{eq11}) should be considered as
more exact than the model (\ref{eq12}). At the same time, if we note
that for $n\leq 1.0\times 10^{17}\,\mbox{cm}^{-3}$ the
conductivity is small, it should be
expected that both models
should lead to practically identical results. This is, however,
not so. In Fig.~5 we present the computational results for the
difference Casimir force using Eq.~(\ref{eq6}). The solid line
$1a$ demonstrates the values of the difference Casimir force versus
separation for the patterned Si plate with a higher concentration of
charge carriers $\tilde{n}_1$ computed under the assumption that
the lower concentration section of the plate is described by Eq.~(\ref{eq12}),
i.e., the conductivity processes at low frequencies are neglected.
The dashed line $1b$ shows the difference Casimir force as a function
of separation computed with the same $\tilde{n}_1$ but taking into
account the conductivity processes at low frequencies in accordance
with Eq.~(\ref{eq11}). As is seen from the comparison of lines
$1a$ and $1b$, the difference Casimir forces computed using
Eqs.~(\ref{eq12}) and (\ref{eq11}) differ by 1.2\,pN at a
separation $z=100\,$nm and this difference slowly decreases to
approximately 0.14\,pN at a separation $z=300\,$nm.
The lines $2a$ and $2b$ present similar results for the case when
the higher charge carrier density is equal to $\tilde{n}_2$.
As is seen from Fig.~5, decreasing the higher concentration
by an order of magnitude decreases the predicted magnitude
of the difference Casimir force by more than two times, but
leaves the same gap between the predictions of two different
models of the permittivity at low frequencies.

Importantly, our predictions do not depend on the discussions
mentioned in the Introduction on the optical properties of metals
at quasi-static frequencies \cite{9,10,11,12,13,14,15,16}.
The resolution of this controversy affects only the value
of the Au reflection coefficient $r_{TE}(0,k_{\bot})$ at zero frequency.
The latter, however, does not contribute to the difference Casimir
force (\ref{eq6}) and pressure (\ref{eq9}) because for dielectrics
and semiconductors $r_{TE;\tilde{n},n}(0,k_{\bot})=0$
regardless of what model (\ref{eq11}) or (\ref{eq12}) is used
for the description of the dielectric permittivity at low
frequencies. The obtained difference between the lines $1a-1b$ and
$2a-2b$ in Fig.~5 is completely explained by the different contributions
of semiconductor reflection coefficient $r_{TM;n}(0,k_{\bot})$ when
one uses Eq.~(\ref{eq11}) or Eq.~(\ref{eq12}) to describe the
dielectric permittivity at low frequencies.
Regarding the semiconductor section
with a higher charge carrier density $\tilde{n}$, from
Eqs.~(\ref{eq8}), (\ref{eq10}) it is always valid that
\begin{equation}
r_{TM;\tilde{n}}(0,k_{\bot})=1.
\label{eq13}
\end{equation}
\noindent
However, for the section of the plate with a lower charge carrier density $n$
it follows from Eq.~(\ref{eq8}) that
\begin{equation}
r_{TM;{n}}(0,k_{\bot})=1\quad\mbox{or}\quad
r_{TM;{n}}(0,k_{\bot})=
\frac{\varepsilon^{Si}(0)-1}{\varepsilon^{Si}(0)+1}
\label{eq14}
\end{equation}
\noindent
when Eq.~(\ref{eq11}) or Eq.~(\ref{eq12}) are used, respectively.
Thus, the difference between the lines $1a$ and $1b$ (and the same
difference between the lines $2a$ and $2b$) can be found
analytically. Taking only the zero-frequency contribution in
Eq.~(\ref{eq6}) and subtracting the difference Casimir force
calculated using Eq.~(\ref{eq11}) [model ($b$)] from
the difference Casimir force
calculated using Eq.~(\ref{eq12}) [model ($a$)] one obtains
\begin{equation}
\Delta F_{a}^{(0)}-\Delta F_{b}^{(0)}=
-\frac{k_BTR}{8z^2}\left\{\zeta(3)-\mbox{Li}_3
\left[\frac{\varepsilon^{Si}(0)-1}{\varepsilon^{Si}(0)+1}
\right]\right\}.
\label{eq15}
\end{equation}
\noindent
Here $\zeta(z)$ is the Riemann zeta function, and Li${}_3(z)$ is
the polylogarithm function. The results using the
analytic Eq.~(\ref{eq15}) coincide with the differences between
the lines $1a-1b$ and $2a-2b$ in Fig.~5 computed
numerically.

In the experiment \cite{24} the difference Casimir force between
Au coated sphere and Si plate illuminated with laser pulses
was first measured. In the presence of light the charge carrier
density was about $2\times 10^{19}\,\mbox{cm}^{-3}$ and in the
absence of light of about $5\times 10^{14}\,\mbox{cm}^{-3}$.
The experimental data were shown to be in agreement with model
($a$) which uses the finite static dielectric permittivity of Si.
The model ($b$) which includes the dc conductivity of Si
was excluded at 95\% confidence within the separation region
from 100 to 200\,nm. As was discussed above, in the framework of
the Lifshitz theory this result is rather unexpected.
Bearing in mind that illumination with laser pulses leads to
several additional sources of errors discussed in Ref.~\cite{24},
it is of vital interest to verify the obtained conclusions in a more
precise experiment with patterned Si plates. The comparison of the
experimental sensitivities presented in Sec.~II with the magnitudes
of the difference Casimir forces computed here using different
theoretical models demonstrate that the proposed experiment with
patterned semiconductor plate will bring decisive results on the
discussed problems in the Lifshitz theory at nonzero temperature.

The calculations of the difference Casimir pressure determined
in the dynamic mode of the proposed experiment leads to results
analogous to those for the difference force. The calculation
results using Eq.~(\ref{eq9}) with the same values of parameters as above
and two models of lower conductivity Si are presented in Fig.~6.
Here the difference Casimir pressures between an Au plate and a
patterned Si plate with the higher densities of charge carriers
$\tilde{n}_{1,2}$ (one section of the plate) and lower $n$ (another
section of the plate) are shown with solid lines $1a$ and
$2a$, respectively,
computed under the assumption that Si with the lower $n$ possesses a
finite permittivity (\ref{eq12}) at zero frequency. The dashed lines
$1b$ and $2b$ are obtained under the assumption that Si with the lower $n$
is described by the permittivity (\ref{eq11}) which goes to infinity
when the frequency goes to zero. As is seen in Fig.~6, the difference
Casimir pressure with a patterned plate with charge carrier
densities $\tilde{n}_1$ and $n$ equals 250\,mPa at a separation
$z=100\,$nm [model ($a$) of low conductivity section of the plate] and
the difference in predictions for the two models equals 38.6\,mPa.
The proposed experiment of the difference Casimir pressure can
reliably discriminate between the solid and dashed lines in
Fig.~6 thus providing one more test for the Lifshitz theory
at nonzero temperature.

Notice that in a similar way to the force, the differences between
the lines $1a-1b$ and $2a-2b$ in Fig.~6 are expressed analytically
by taking the zero-frequency contributions in Eq.~(\ref{eq9}):
\begin{equation}
\Delta P_{a}^{(0)}-\Delta P_{b}^{(0)}=
-\frac{k_BT}{8\pi z^3}\left\{\zeta(3)-\mbox{Li}_3
\left[\frac{\varepsilon^{Si}(0)-1}{\varepsilon^{Si}(0)+1}
\right]\right\}.
\label{eq16}
\end{equation}
\noindent
Calculations using Eq.~(\ref{eq16}) lead to the same differences between
the lines $1a-1b$ and $2a-2b$ as were computed numerically in Fig.~6.

\section{Proposed experiment on the modulation of the Casimir
force through an insulator-metal transition}

The exciting possibility for the modulation of the Casimir force due
to a change of charge carrier density is offered by semiconductor
materials that undergo the insulator-metal transition with the
increase of temperature. Such a transition leads to a change of
the carrier density of order $10^4$. Although in literature it is
common to speak about insulator-metal transition, this
can be considered as a
transformation between two semiconductor phases with lower and
higher charge carrier densities $n$ and $\tilde{n}$, respectively.
As was shown above, this is sufficient to bring about a large
change in the Casimir force. From a fundamental point of view,
the modulation of the Casimir force due to the phase transition
will offer one more precision test of the role of conductivity
and optical properties in the Lifshitz theory of the Casimir force.
This experiment suggests some advantages as compared to the difference
force measurement with a patterned plate considered in Secs.~II and III.
First, because of the large change in the magnitude and bandwidth of
the optical properties in a phase transition, the modulation of the
Casimir force will be larger. Second, an insulator-metal
transition does not require the special fabrication of patterned plates
with one section having a high carrier density, which might not be
compatible with robust device design. Keeping in mind that the increase of
temperature necessary for the phase transition can be induced by laser light,
this opens up the possibility of radically new nanomechanical devices
using the Casimir force in image detection. The phase transition can be
also brought about through electrical heating of the
material.

In this experiment we propose to measure the change of the Casimir
force acting between an Au coated sphere and a vanadium dioxide
(VO${}_2$) film deposited on sapphire substrate
which undergoes the insulator-metal transition with the increase of temperature.
It has been known that VO${}_2$ crystals and thin films undergo an abrupt
transition from semiconducting monoclinic phase at room temperature
to a metallic tetragonal phase at 68\,${}^{\circ}$C
\cite{25,39,40,41,42,43,44}. The phase transition causes the
resistivity of the sample to decrease by a factor of $10^4$
from 10\,$\Omega\,$cm to $10^{-3}\,\Omega\,$cm (i.e., the same change as
for two semiconductor half plates in Sec.~II with lower and higher
charge carrier densities).  In addition, the optical transmission
for a wide region of wavelengths extending from $1\,\mu$m to
greater than $10\,\mu$m, decreases by more than a factor of 10--100.

The schematic of the experimental setup is shown in Fig.~7. In
this figure, light from a chopped 980\,nm laser will be used to
heat the VO${}_2$ film \cite{39,40}. About 10--100\,mW power
of the 980\,nm laser is required to bring about all optical
switching of VO${}_2$ films. The same procedure as outlined in Sec.~II
(the static technique) will be used in the measurement of the
modulation of the Casimir force including the interferometric
detection of cantilever deflection. The schematic of the setup
is similar to the one used in Ref.~\cite{24} in the demonstration
of optically modulated dispersion forces. An important point is
that in Ref.~\cite{24} the absorption of light from
a 514\,nm Ar laser led to an increase of charge carrier density.
By contrast, here the wavelength of a laser is selected in such a way
that light only leads to heating of a VO${}_2$ film but does not change
the number of free charge carriers \cite{39}.

As a first step towards studying the role of the insulator-metal
transition in the Casimir force, we have recently fabricated
thin films of VO${}_2$ on sapphire plates. The preliminary
results are shown in Fig.~8. It is observed that we have obtained
more than a factor of 10 change in the resistivity of the film.
These films were prepared by thermal evaporation of VO${}_2$ powder.
While films of appropriate thickness approaching 100\,nm and roughness of
about 2\,nm (shown in Fig.~9) can be obtained by this procedure,
it is not optimal as it leads to the non-stoichimetric formation of
the mixed valence states of the vanadium oxide (VO${}_x$).
In the future rf magnetron sputtering will be used to make the
films \cite{39}. VO${}_2$ films using this technique have been shown
to have the $10^4$ change in resistivity and a corresponding large
change in optical reflectivity and spectrum.

The aim of the proposed experiment on the influence of insulator-metal
transition on the Casimir force is two fold: applications for
actuation of nanodevices through a modulation of the Casimir force,
and to perform fundamental tests on the theory of dispersion forces.
To accomplish this, two types of measurements are planned.
In the first we plan to demonstrate the modulation of the Casimir
force through an optical switching of the insulator-metal
transition. This modulation will lead to novel microdevices as
optical and electrical switches, optical modulators, optical filters
and IR detectors that can be actuated optically through the
absorption of IR radiation. Importantly, such devices can be
integrated with Si technology which is used in the fabriaction of
microelectromechanical systems \cite{39}.
In the second type of measurements, the variable temperature
atomic force microscope described in Sec.~II will be used to
perform precision measurements of the Casimir force between a gold coated
sphere and VO${}_2$ film. Here, the Casimir force will be measured
at different temperatures from room temperature through 80\,${}^{\circ}$C.
This temperature range spans the dielectric (semiconducting) and metal
regions of VO${}_2$. Careful comparison of the experimental data and the
theory (see the next section) will be done to understand the role of
conductivity and losses in both phases of VO${}_2$.

\section{Calculation of the Casimir
force in an insulator-metal transition}

The Casimir force acting between a metal coated sphere and the VO${}_2$ film
on a sapphire plate
both before and after the phase transition (i.e., in the insulating and
metal phases or, more exactly, in the semiconductor phases with
lower and higher charge carrier densities) are expressed by the
Lifshitz formulas in accordance with Eqs.~(\ref{eq1}) and (\ref{eq6}).
As above, we label the higher concentration of charge carriers
$\tilde{n}$ and the lower concentration $n$. To compute the Casimir
force before and after the phase transition one needs the optical properties
of VO${}_2$ on a sapphire plate in a wide frequency region.

In Ref.~\cite{43} the dielectric permittivity of VO${}_2$ is
measured and fitted to the oscillator model
for both bulk VO${}_2$ and for a system of 100\,nm thick VO${}_2$ film
deposited on bulk sapphire plate within the frequency
region from 0.25\,eV to 5\,eV.
This modelling was performed both before and after the phase transition.
Typical thickness of sapphire substrate is of about 0.3\,mm, i.e.,
the same as the thickness of patterned Si plate in Secs.~II and III.
Because of this, when calculating the Casimir force between gold coated
sphere and VO${}_2$ film on sapphire substrate, we can use the Lifshitz
formala for bulk test bodies (see Sec.~III for details).
The application region of the models presented in Ref.~\cite{43}
should be extended in order to perform
computations of the Casimir force within the separation region from
100 to 300\,nm where contribution from
 optical data up to about 10\,eV have to be taken into account.
For this purpose, we have supplemented equations of Ref.~\cite{43} with
additional terms taking into account the frequency-dependent
electronic transitions at high frequencies \cite{45,46}.
As a result, the effective dielectric permittivity of the VO${}_2$ film
on a sapphire substrate before
the phase transition (at $T=300\,$K) is given by
\begin{equation}
\varepsilon_n(i\xi_l)=1+\sum\limits_{i=1}^{7}
\frac{s_{n,i}}{1+\frac{\xi_l^2}{\omega_{n,i}^2}+
\Gamma_{n,i}\frac{\xi_l}{\omega_{n,i}}}+
\frac{\varepsilon_{\infty}^{(n)}-1}{1+
\frac{\xi_l^2}{\omega_{\infty}^2}}.
\label{eq17}
\end{equation}
\noindent
Here the values of the oscillator frequencies $\omega_{n,i}$,
dimensionless relaxation parameters $\Gamma_{n,i}$ and of the
oscillator strengths $s_{n,i}$ taken from Fig.~5 in Ref.~\cite{43}
are presented in Table~I.
The constants related to the contribution of high-frequency electronic
transitions [the last term on the right-hand side of
Eq.~(\ref{eq17})] are $\varepsilon_{\infty}^{(n)}=4.26$ \cite{43}
and $\omega_{\infty}=15\,$eV. If we put $\xi_l=0$ in the last
term on the right-hand side of Eq.~(\ref{eq17}), this equation
is the same as the result in Ref.~\cite{43}.

After the phase transition we have a phase with increased charge
carrier density $\tilde{n}$. Similar to Eq.~(\ref{eq10}) the effective
dielectric permittivity of the VO${}_2$ film on a sapphire substrate
can be described by the dielectric permittivity
\begin{eqnarray}
&&
\varepsilon_{\tilde{n}}(i\xi_l)=1+
\frac{\omega_{p;\tilde{n}}^2}{\xi_l(\xi_l+\gamma_{\tilde{n}})}
\label{eq18} \\
&&
\phantom{aa}+
\sum\limits_{i=1}^{4}
\frac{s_{\tilde{n},i}}{1+\frac{\xi_l^2}{\omega_{\tilde{n},i}^2}+
\Gamma_{\tilde{n},i}\frac{\xi_l}{\omega_{\tilde{n},i}}}+
\frac{\varepsilon_{\infty}^{(\tilde{n})}-1}{1+
\frac{\xi_l^2}{\omega_{\infty}^2}}.
\nonumber
\end{eqnarray}
\noindent
Parameters $\omega_{\tilde{n},i}$, $\Gamma_{\tilde{n},i}$ and
$s_{\tilde{n},i}$ can be found in Fig.~6 of Ref.~\cite{43} and are listed in
Table~II. The other parameters are
$\varepsilon_{\infty}^{(\tilde{n})}=3.95,$
$\omega_{p;\tilde{n}}=3.33\,$eV,
$\gamma_{\tilde{n}}=0.66\,$eV \cite{43}. Setting $\xi_l=0$ in
the last term on the right-hand side of Eq.~(\ref{eq18})
(this term takes high-frequency electronic transitions into account), returns
(\ref{eq18}) to the original form suggested in Ref.~\cite{43}.
Note that the recently suggested model for the dielectric
permittivity of VO${}_2$ films \cite{47} is applicable not only
before and after a phase transition but also at intermediate
temperatures. This model is, however, restricted to a more
narrow frequency region from 0.73 to 3.1\,eV and uses the
simplified description of two oscillators before the phase
transition and only one oscillator with nonzero frequency
after it.

In Fig.~10, the effective dielectric permittivity of VO${}_2$ film of 100\,nm
thickness on sapphire substrate before and after the phase transition, as given in
Eqs.~(\ref{eq17}) and (\ref{eq18}) is shown by the solid lines
1 and 2, respectively. In the same figure, the dielectric permittivity
of Au versus frequency is shown as dots. The vertical line indicates the
position of the first Matsubara frequency at $T=340\,$K (i.e., in the
region of the phase transition).

In Fig.~11 we present the computational results for the Casimir force
between the Au coated sphere and VO${}_2$ film on sapphire substrate
versus separation
obtained by the substitution of the dielectric permittivity (\ref{eq17})
(VO${}_2$ before the phase transition in solid line 1) and (\ref{eq18})
(VO${}_2$ after the phase transition in solid line 2) into the
Lifshitz formula. As is seen in Fig.~11, after the phase transition the
magnitudes of the Casimir force increase due to an increase
in the charge carrier
density. For a comparison with the proposed
experiment on the difference Casimir
force from a patterned Si plate, in Fig.~12 (solid line) we plot the
difference of the Casimir forces after and before the phase
transition, i.e., the difference of lines 2 and 1 in Fig.~11. It is
seen that the difference Casimir force from a phase transition
changes from 13\,pN at $z=100\,$nm to 1.2\,pN at $z=300\,$nm, i.e.,
the magnitudes of the difference from the phase transition are
greater than that from the patterned Si plate.

The difference Casimir force in the insulator-metal phase
transition provides us with one more test on the proper
modelling of the dielectric permittivity in the Lifshitz theory
of dispersion forces. Similar to Sec.~III, we arrive at different
results for the difference Casimir force
after and before the phase transition if the conductivity of
a dielectric VO${}_2$ at zero frequency is taken into account
in our computations. The shift in the values of the difference
Casimir force is completely determined by the change of the
zero-frequency term in the Lifshitz formula. By analogy with
Eq.~(\ref{eq15}) it follows:
\begin{equation}
\Delta F_{a}^{(0)}-\Delta F_{b}^{(0)}=
-\frac{k_BTR}{8z^2}\left\{\zeta(3)-\mbox{Li}_3\left[
\frac{\varepsilon^{{\rm VO}_{2}}(0)-1}{\varepsilon^{{\rm VO}_2}(0)+1}
\right]\right\},
\label{eq19}
\end{equation}
\noindent
where $b$ represents the case when the dc conductivity of an
insulating VO${}_2$ is taken into account, and $a$
represents the case when
insulating VO${}_2$ is described by the permittivity (\ref{eq17}).
From Eq.~(\ref{eq17}) and Table~I one obtains
\begin{equation}
\varepsilon^{{\rm VO}_{2}}(0)\equiv\varepsilon_n(0)=
\varepsilon_{\infty}^{(n)}+\sum\limits_{i=1}^{7}s_{n,i}
=9.909.
\label{eq20}
\end{equation}

In Fig.~12 the difference Casimir force between an Au coated sphere
and VO${}_2$ film on sapphire substrate
after and before the phase transition computed
including the dc conductivity of insulating VO${}_2$ versus separation
is plotted with the dashed line. The difference between the solid and
dashed lines is determined by Eq.~(\ref{eq19}). This difference
changes from 1.6\,pN at $z=100\,$nm to 0.2\,pN at $z=300\,$nm.
Thus, in the phase transition experiment the predicted discrepances
between the two theoretical approaches to the description of
conductivity properties at low frequencies are larger than in the
experiment with the patterned semiconductor plate. This will help
to experimentally discriminate between the two approaches and
deeply probe the role of the material properties in the Lifshitz
theory at nonzero temperature.

\section{Conclusions and discussion}

In the above we have proposed two experiments on the measurement
of the difference Casimir force acting between a metal coated
sphere and a semiconductor with different charge carrier
densities. One of these experiments is based on the formation of
a special patterned Si plate, two sections of which have charge
carrier densities differing by several orders of magnitude.
The measurement scheme in this experiment is differential, i.e.,
adapted for the direct measurement of the difference in the
Casimir forces between the sphere and each section of the patterned
plate. This allows one to obtain high precision within a wide
measurement range. Using the dynamic measurement technique,
this experiment also permits the measurement of the difference
Casimir pressure between two parallel plates one of which is coated
with gold and the other is patterned and consists of two sections
with different charge carrier densities.

Another proposed experiment directed to the same objective is novel
and uses the insulator-metal phase transition
brought about by an increase
of temperature in the measurements of the Casimir force. This
transition also leads to the change of charge carrier density
by several orders of magnitude while not requiring the formation
of special patterned samples. The expected difference in the
Casimir forces after and before the phase transition is even
larger than in the experiment with the patterned Si plate.

Both proposed experiments are motivated by the uncertainties
in the application of the theory of dispersion forces at nonzero
temperature. As was shown above, different models of the
conductivity of semiconductors at low frequencies used
in the literature predict variations of the difference Casimir
force at the level of 1\,pN.
An even greater concern is that the model taking into account
the dc conductivity of dielectrics violates the Nernst heat
theorem \cite{49a,49b,49c}.
We have reported an apparatus
developed at UC Riverside that has the sensitivity of force
measurements on the level of 1\,fN and is well adapted
for the systematic investigation of the proposed effects in
a wide range of separations. This apparatus includes an
interferometer based atomic force microscope operated in
high vacuum over a temperature range from 360\,K to 4\,K.
The proposed experiments are feasible using the developed
techniques and will aid in the resolution of theoretical
problems on the application of the Lifshitz theory at
nonzero temperature to real materials.

Another motivation of the proposed experiments is in the
application to nanotechnology. The separations between the
adjacent surfaces in micro- and nanoelectromechanical
devices are rapidly falling to a region below a
micrometer where the Casimir force becomes dominant.
Keeping in mind that semiconducting materials are
used for micromachines, the detailed investigation
of the dependence
of the Casimir force on the properties of semiconductors is
important. The proposed experiments and related theory
clearly demonstrate that it is possible to control the
Casimir force with semiconductor surfaces by changing the
charge carrier density with doping or excitation. This opens
new opportunities discussed above for using the Casimir force
in both the operation and function of  novel
nanomechanical devices.

In addition to the previously performed experiments on the
Casimir force (see review \cite{4} and
Refs.~\cite{5,6,7,8,20,22,23,24}) currently a number
of new experiments have been proposed in the literature.
Thus, Ref.~\cite{48} proposes to measure the Casimir torque
between two parallel birefringent plates with in-plane
optical anisotropy separated by either a vacuum or ethanol.
In Ref.~\cite{48a} it is suggested to measure the vacuum
torque between corrugated mirrors.
References \cite{49,50,51} propose the measurements of the
Casimir force between metallic surfaces at large
separations of a few micrometers as a spherical lens and
a plate, a cylinder and a plate or two parallel plates.
These experiments are aimed at resolving the
theoretical problems arising in the Lifshitz theory when it
is applied to real metals. In Refs.~\cite{52,52a} a
proposal to measure the influence of the Casimir energy
on the value of the critical magnetic field in the
transition from a superconductor to a normal state
has been made.
References~\cite{53,53a} proposed the measurement of the dynamic
Casimir effect resulting in the creation of photons.
The experiments proposed here on the difference Casimir
force through the use of patterned semiconductor samples
or using the insulator-metal phase transition indicate important
new promising directions for future
investigations in the Casimir effect.

\section*{Acknowledgments}

G.L.K.and V.M.M. are grateful to the Department of Physics and
Astronomy of the University of California (Riverside)
for its kind hospitality.
The work on the difference Casimir force with patterned
semiconductor samples and insulator-metal phase
transition was supported by the
DOE Grant No.~DE-FG02-04ER46131.
The development of the interferometer based low temperature
AFM was supported by the NSF Grant No.~PHY0355092.
R.C.-G. is grateful for the financial support of UCMEXUS
and CONACYT.

\newpage
{\bf Figures} \hfill \\[2mm]
{\bf Fig.~1.} {(Color online)
Schematic diagram of the experimental setup for the
measurement of the difference Casimir force.
The patterned Si plate with two sections of different
dopant densities is mounted on a piezo below the Au
coated sphere attached to a cantilever of an atomic
force microscope. The piezo oscillates in the horizontal
direction above different regions of the plate causing the
flexing of the cantilever in response to the
Casimir force.
}
{\bf Fig.~2.} {(Color online)
Steps in the fabrication of the patterned Si plate with
patterned doping (see text for more details).
}
{\bf Fig.~3.} {(Color online)
Image of the interferometer based variable temperature atomic
force microscope with the force sensitivity up to 1\,fN
fabricated at UC, Riverside. The critical components are labeled.
}
{\bf Fig.~4.} {
The dielectric permittivity of Si along the imaginary
frequency axis for samples with high concentration of charge
carriers $\tilde{n}_1$ and $\tilde{n}_2$ is shown by the
solid lines 1 and 2, respectively. For the sample with
a low concentration of charge carriers $n$ the permittivity
versus frequency is shown by dashed lines $a$ and $b$ based
on whether the static permittivity is finite or infinitely
large. The premittivity of Au is indicated by the dotted line.
}
{\bf Fig.~5.} {
The difference Casimir forces versus separation
in the case when the higher concentration of charge carriers
is equal to $\tilde{n}_1$ and the sample with a lower
concentration, $n$, is described by a finite or infinitely
large static permittivity are shown by the solid line 1$a$
and dashed line 1$b$, respectively.
The analogous difference forces
when the higher concentration of charge carriers
is equal to $\tilde{n}_2$ are shown by the solid line 2$a$
and dashed line 2$b$.
}
{\bf Fig.~6.} {
The difference Casimir pressures versus separation
in the case when the higher concentration of charge carriers
is equal to $\tilde{n}_1$ and the sample with a lower
concentration, $n$, is described by a finite or infinitely
large static permittivity are shown by the solid line 1$a$
and dashed line 1$b$, respectively.
The analogous difference pressures
when the higher concentration of charge carriers
is equal to $\tilde{n}_2$ are shown by the solid line 2$a$
and dashed line 2$b$.
}
{\bf Fig.~7.} {
Schematic of the experimental setup for the observation
of modulation of the Casimir force in an insulator-metal
phase transition. Light from a chopped 980\,nm laser
heats a VO${}_2$ film leading to a phase transition to
a state with higher concentration of charge carriers
(sapphire substrate is not shown).
Cooling in between pulses causes the transition to a state
with lower concentration of carriers. The cantilever
of an atomic force microscope flexes in response to the
difference Casimir force.
}
{\bf Fig.~8.} {(Color online)
Preliminary results on the resistance of VO${}_2$ film grown at
UC Riverside as a function of temperature are shown as black
squares (heating) and dots (cooling).
}
{\bf Fig.~9.} {(Color online)
Morphology of the same VO${}_2$ film, as in Fig.~8, grown
by thermal evaporation. The heights of roughness peaks are
of about 2\,nm.
}
{\bf Fig.~10.} {
The effective dielectric permittivity of VO${}_2$ film
on sapphire substrate along the
imaginary frequency axis before and after the phase
transition are shown by the solid lines 1 and 2,
respectively. The permittivity of Au is indicated by the
dotted line.
}
{\bf Fig.~11.} {
The Casimir force between an Au coated sphere and VO${}_2$
film on sapphire substrate
versus separation before and after the phase transition
are shown by the solid lines 1 and 2, respectively.
}
{\bf Fig.~12.} {
The difference of the Casimir forces after and before the phase
transition versus separation computed using a finite static
dielectric permittivity (solid line) and taking into account the
dc conductivity of VO${}_2$ in a dielectric state (dashed line).
}

\begingroup
\squeezetable
\begin{table}
\caption{Values of the oscillator resonant frequencies
$\omega_{n,i}$, dimensionless relaxation parameters
$\Gamma_{n,i}$ and oscillator strengths $s_{n,i}$ of
VO${}_2$ film on sapphire substrate before the phase transition.
}
\begin{ruledtabular}
\begin{tabular}{llll}
i & $\omega_{n,i}\,$(eV) & $\Gamma_{n,i}$ & $s_{n,i}$  \\
\hline
1 & 1.02 & 0.55 & 0.79 \\
2 & 1.30 & 0.55 & 0.474 \\
3 & 1.50 & 0.50 & 0.483 \\
4 & 2.75 & 0.22 & 0.536 \\
5 &3.49 & 0.47 & 1.316 \\
6 & 3.76 & 0.38 & 1.060 \\
7 & 5.1 & 0.385 & 0.99
\end{tabular}
\end{ruledtabular}
\end{table}
\endgroup
\begingroup
\squeezetable
\begin{table}
\caption{Values of the oscillator resonant frequencies
$\omega_{\tilde{n},i}$, dimensionless relaxation parameters
$\Gamma_{\tilde{n},i}$ and oscillator strengths
$s_{\tilde{n},i}$ of
VO${}_2$ film on sapphire substrate after the phase transition.}
\begin{ruledtabular}
\begin{tabular}{llll}
i & $\omega_{\tilde{n},i}\,(eV)$ & $\Gamma_{\tilde{n},i}$
& $s_{\tilde{n}n,i}$  \\
\hline
1 & 0.86 & 0.95 & 1.816 \\
2 & 2.8 & 0.23 & 0.972 \\
3 & 3.48 & 0.28 & 1.04 \\
4 & 4.6 & 0.34 & 1.05
\end{tabular}
\end{ruledtabular}
\end{table}
\endgroup

\end{document}